# Non-volatile rewritable frequency tuning of a nanoelectromechanical resonator using photoinduced doping

*David J. Miller[1,2,3], Andrew Blaikie[1,2,3], and Benjamín J. Alemán[1,2,3,4,*]*

[1]Department of Physics, University of Oregon, Eugene, Oregon, 97403, United States

[2]Materials Science Institute, University of Oregon, Eugene, Oregon, 97403, United States

[3]Center for Optical, Molecular, and Quantum Science, University of Oregon, Eugene, Oregon, 97403, United States

[4]Phil and Penny Knight Campus for Accelerating Scientific Impact, University of Oregon, Eugene, Oregon, 97403, United States

ABSTRACT:

Arrays of nanoelectromechanical resonators (NEMS) have shown promise for a suite of applications, ranging from nanomechanical information processing technologies to high-resolution mass spectroscopy systems. In these arrays, the mechanical resonance frequency can form the base unit of information, like the voltage level in analog electronics. A fundamental challenge towards broader adoption of NEMS arrays is a lack of viable frequency tuning methods, which must simultaneously allow for persistent and reversible control of single resonators while also being scalable to large arrays of devices. In this work, we demonstrate an electro-optic tuning method for graphene-based NEMS that meets these needs. Our method uses a focused laser to locally photoionize defects on an individual resonator. After the laser is turned off, the trapped charge created by photoionization persists and applies spatially localized electrostatic strain to the resonator, thereby tuning its frequency. Our approach has a persistence time of several days and can repeatedly write and erase the state of a single device with a high degree of precision. We show the scalability of this technique by aligning the frequencies of spatially separated NEMS devices and discuss potential implications of this tuning method when applied to both single devices and to programmable NEMS networks.

MAIN TEXT:

Nanoelectromechanical (NEMS) resonators are among the most sensitive detectors of mass[1], charge[2], and force[3], and have enabled discoveries of nonlinear phenomena[4] as well as tests of macroscopic quantum mechanics[5]. Much like their electronic device counterparts, the potential of NEMS grows when they are built up into large-scale arrays and networks. These arrays have already enabled neutral mass spectroscopy[6], and have been proposed for ultralow-power alternatives to traditional analog electronics[7] as well as nanomechanical information technologies like memory[8,9], logic[10], and computing[11,12]. Moreover, programmable arrays, where the frequency and coupling of each individual resonator are precisely controlled, are expected to enable tunable acoustic bandgaps[13] and lensing[14], and more exotic systems like topological materials[15], metamaterials[16], and as a potential platform for neuromorphic computing[17,18] or simulating complex networks[19].

To realize the promise of NEMS arrays, methods are needed to efficiently program and interface with large networks of NEMS. Tuning the resonance frequency of individual NEMS offers one such method but the frequency tuning must be simultaneously persistent (*i.e.* have memory), reversible, fast, and operate over a large frequency range. There have been numerous demonstrations of NEMS tuning methods[20], but each is accompanied by significant drawbacks and challenges when applied to large arrays of devices. Active tuning methods, such as electrostatic gating[21] or local heating[22], are reversible and can achieve a large tuning range, but are inherently transient. Thus, to maintain the tuned frequency, active tuning requires a continuous, separate external force for each NEMS resonator, making them impractical for integration into large arrays. In contrast, passive methods permanently modify the NEMS structure, for instance by adding or removing mass[23–25], can achieve persistent tuning but at the expense of rewritability. Several NEMS tuning approaches have attempted to combine persistence and reversibility, including mass electromigration along suspended carbon nanotubes[26] and etching/depositing of mass with a focused ion beam[27], but these techniques require an electron microscope and *in-situ* nanomanipulation, which severely impedes practicality and scalability. Moreover, these tuning schemes suffer from poor frequency resolution, a limited tuning range (∼10%), slow speed, and limited cyclability.

Here, we demonstrate a non-volatile and rewritable frequency tuning method for graphene-based two-dimensional (2D) NEMS[21,24,28–31]. In our approach, we use a focused laser and two global electrical contacts to create locally photoionized defects[32–36] on an individual resonator. After the optical and electrostatic fields are removed, the trapped charge created by the photoionization persists and applies spatially localized electrostatic strain to the resonator, thereby tensioning the resonator and tuning its frequency. Our approach is robustly rewritable over a large tuning range and persistent over many days with no need for external power, and is also exceptionally fast and fully scalable to NEMS arrays of arbitrary size. By providing a facile means to locally address the frequency of a NEMS resonator, this work lays the groundwork for fully programmable large-scale NEMS lattices and networks[16,37].

Our frequency tuning method relies on principles similar to electrostatic gate-tuning[21,24], which is among the most commonly used active tuning methods. In a typical gate-tuning configuration, a NEMS device is suspended above a gate electrode to form a simple parallel-plate capacitor. An electrostatic potential will tension the membrane and tune the resonance frequency to a value $f_0(V_g) \propto \frac{1}{2}\frac{dC_g}{dx}\left(V_g - V_{mCNP}\right)^2$, where $V_g$ is the external gate voltage and $V_{mCNP}$ is called the mechanical charge neutrality point, which is analogous to the electronic charge neutrality point[38]. Non-zero values of $V_{mCNP}$, have routinely been observed in 2D NEMS[24,30] and are typically ascribed to the presence of charged defects. In this case, the effective voltage experienced by the membrane is $V_{eff} = V_g - V_{mCNP}$. Although tuning $f_0$ with an external $V_g$ is commonplace, tuning $f_0$ by controllably modifying $V_{mCNP}$ has not previously been explored and forms the basis of this work.

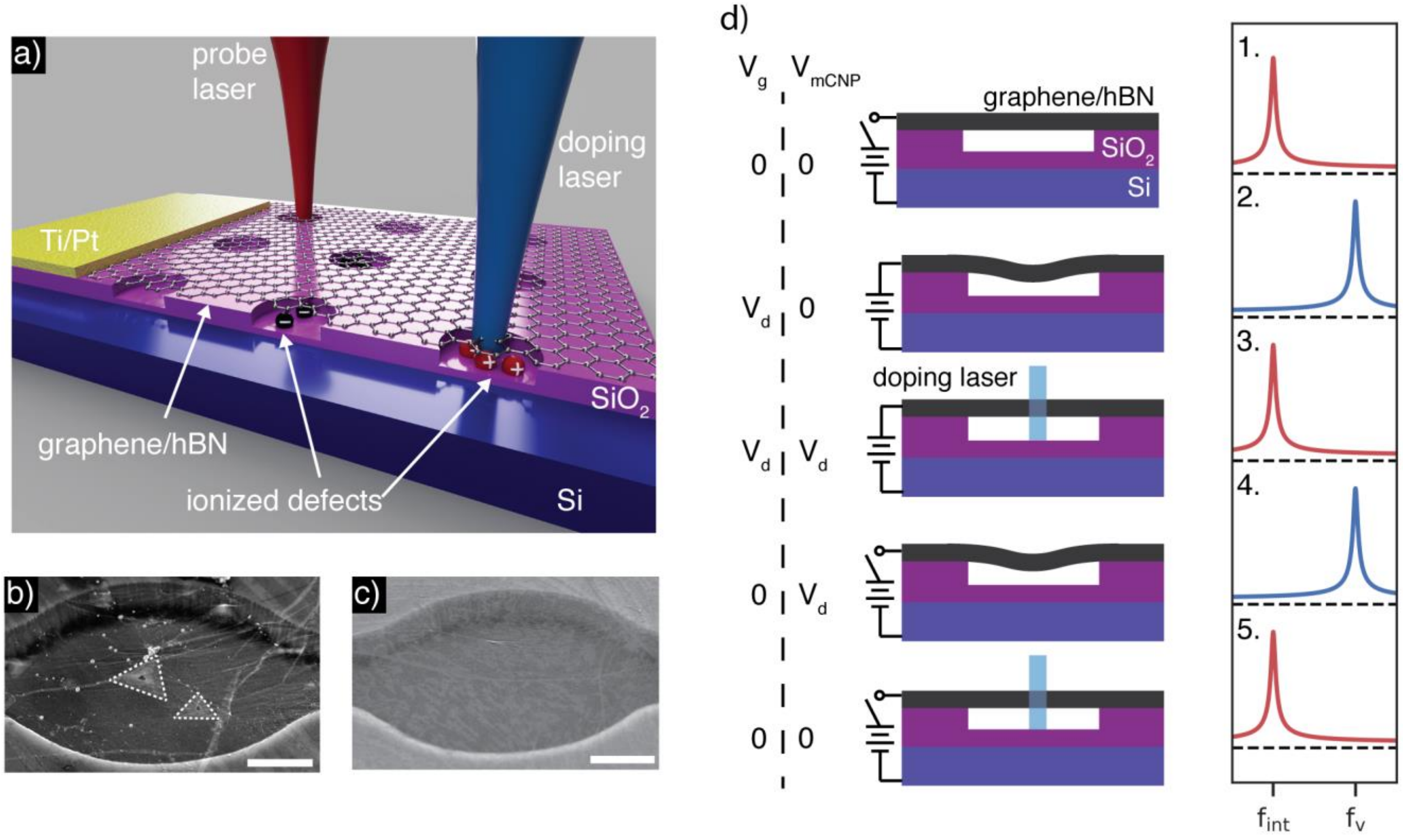

**Figure 1: a)** Schematic of the phototuning effect in 2D NEMS. **b)** SEM image of a gr/hBN drumhead with venting trenches on the sides. Small multilayer islands of h-BN are visible on the surface. Scale = 1 μm. **c)** SEM image of a graphene drumhead. Scale = 1 μm. **d)** The frequency tuning sequence. 1: The device is initially in its intrinsic state with $V_g = 0$ V, $V_{mCNP} = 0$, and resonance frequency $f_{int} = f_0(V_{eff} = 0)$. 2: The device is then biased with $V_g = V_d$, tuning the resonance frequency to $f_0(V_d)$. 3: Photodoping the device while $V_g = V_d$ sets $V_{mCNP}$ to $V_d$ returns the resonance frequency to $f_{int}$. 4: With the gate off, the device is in its phototuned state with a resonance frequency $f_V = f_0(-V_{mCNP})$. 5: Photodoping with $V_g = 0$ now returns the device to its initial configuration, with a resonance frequency $f_{int}$.

We tune $V_{mCNP}$ of an individual 2D NEMS device with spatially resolved photodoping[32–36]. In previous studies of photodoping in graphene, graphene is separated from a global gate-electrode by a stack of dielectrics with varying bandgaps[34]. Simultaneous application of a gate-voltage and a focused laser causes ionized defects to accumulate at the laser focus, shifting the electronic charge neutrality point until the applied gate voltage is neutralized. Crucially, the charge remains trapped after the laser and gate are removed, allowing for intricate, doping patterns to be patterned into the heterostructure. By applying the same principle to suspended graphene NEMS, we can controllably modify $V_{mCNP}$ using only a focused laser and a single global gate voltage $V_g$, allowing us to tune $f_0$ in an arbitrarily large array of devices.

We study this tuning method in NEMS membranes made from both CVD-grown monolayer graphene and a graphene/hBN heterostructure (gr/hBN). Our devices consist of the two-dimensional sheet suspended over $\sim 4-5$ µm diameter circular cavities etched into $SiO_2$ on top of a degenerately doped silicon gate electrode (see Figure 1a-c). A layer of $SiO_2$ ($\sim 300$ nm thick) is left at the bottom of the cavities to prevent shorting and create potential charge traps[33]. The devices are driven with standard electrostatic actuation techniques[21] using the silicon back-gate and a Ti/Pt top contact, and measured using an interferometer operating at $633$ nm[28,31]. Photodoping is typically performed with a power-stabilized $445$ nm diode laser except where noted otherwise.

To set or change the frequency of an individual membrane through photodoping—a process we call phototuning—we apply a bias to a global back-gate (in this case, degenerately doped silicon) while focusing a laser onto the individual, suspended membrane of interest (see Figure 1d). Prior to any photodoping and assuming an initial $V_{mCNP}$ of 0 V, the resonator will be at its intrinsic resonance frequency $f_{int} = f_0(0)$ before the laser or bias are turned on (*step 1*). Then, we set the gate voltage to a value $V_d$, tensioning the membrane and blue shifting the resonance frequency (*step 2*) from $f_{int}$ to $f_0(V_d)$. Next, with the bias still at $V_d$, we turn the laser on to start the charge-trapping process. This brings $V_{mCNP}$ towards $V_d$, lowering $V_{eff}$, and red shifting the resonance frequency. Given enough laser dose, $V_{mCNP}$ saturates at $V_d$ and the frequency returns to $f_{int}$ (*step 3*). After turning the laser and bias off, the frequency immediately blue shifts to $f_0(-V_d)$ (*step 4*), which is the same as $f_0(V_d)$ due to the symmetry of $f_0(V_{eff})$. The phototuned frequency obtained after *step 4* does not require an external gate bias to maintain and is denoted $f_V = f_0(-V_{mCNP})$. Steps 1-4 complete the phototuning "write" function. The frequency can be reset back to $f_{int}$—or "erased"—by zeroing the bias voltage ($V_d = 0$ V) and illuminating the membrane with the laser (*step 5*). We note that the description above represents the ideal case of phototuning. For most devices we study, $V_{mCNP}$ is initially slightly offset from zero and $V_{mCNP}$ saturates at a slightly different value than $V_d$, but neither of these factors affect the crucial properties of the phototuning method.

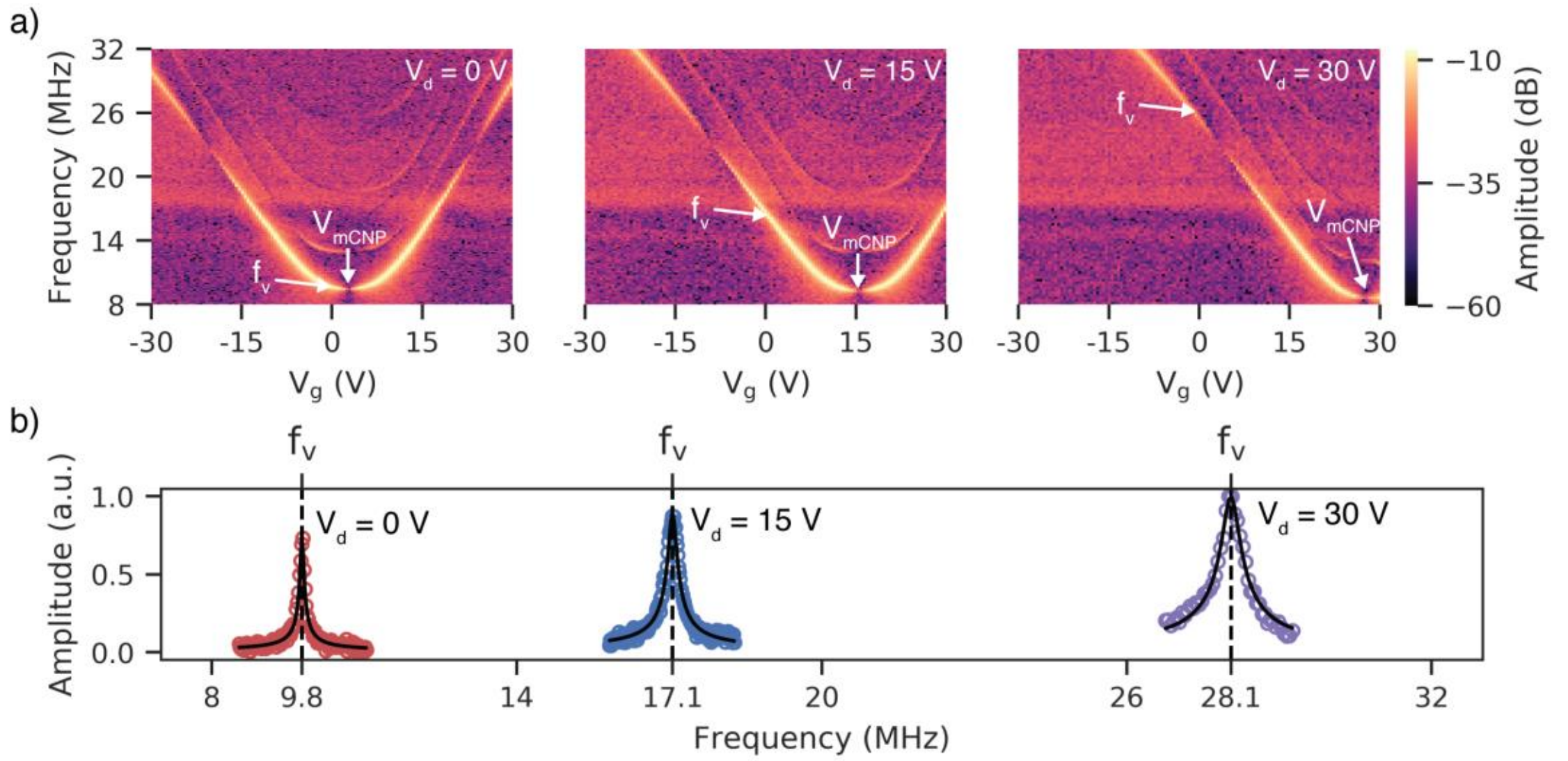

**Figure 2: a)** Electrostatic tuning curves for a gr/hBN drumhead resonator as it is phototuned from the erased state, *i.e.* with $V_d = 0$ V (left), to phototuned states with $V_d = 15$ V (middle) and $V_d = 30$ V (right). The $V_{mCNP}$ is initially located at $\sim 1$ V and shifts to $V_{mCNP} \sim 28$ V after phototuning at $V_d = 30$ V. **b)** Resonant response at $V_g = 0$ V for the three different states in (a). $f_V$ is the center frequency in each of these curves. The resonance frequency increases by $\sim 200\%$ and the $Q$-factor decreases from 135 to 74 between the $V_d = 0$ V and the $V_d = 30$ curves.

The fundamental effects of phototuning are revealed in the gate voltage tuning curves and frequency response spectra. We show gate curves for a gr/hBN device in the erased state ($V_d = 0$ V) and two tuned states ($V_d = 15, 30$ V), in Figure 2a. For these measurements, the membrane was photodoped to saturation by rastering the laser over the area of the drumhead at relatively high power ($\sim 1$ mW/μm$^2$). The fundamental mode is the easiest to resolve, but several higher order modes are also present. The curve shapes are consistent with an electrostatically biased membrane[21,24]. In the erased state, $V_{mCNP}$ is offset from zero by $\sim 1$ V, which is common and indicates the presence of static charged contaminants. With $V_d = 30$ V, $V_{mCNP}$ saturates to $\sim 28$ V, where the new potential arises from charged defects in the h-BN[32,34] and the oxide[33]. Although $V_{mCNP}$ differs by $\sim 27$ V between these states, the gate-dependence of each mode relative to $V_{mCNP}$ doesn't change. Thus, apart from the $V_{mCNP}$ shift, the phototuning process does not alter the mechanical characteristics of the device in a significant way, unlike most passive tuning methods[22,25].

The individual resonance curves corresponding to the three states are shown in Figure 2b. As $V_{mCNP}$ is tuned from ~1 V to ~28 V, $f_V$ increases from ~ 9.8 MHz to ~ 28.1 MHz, a change of ~ 200%. The $Q$ of the 28.1 MHz peak ($Q = 74$) is smaller than that of the 9.8 MHz peak ($Q = 135$), just as in the case of an applied electrostatic backgate[29]. Although the central focus of this work is to report the phototuning effect, it also reveals that measuring or driving[21] 2D NEMS with an optical probe can lead to unintended frequency drift.

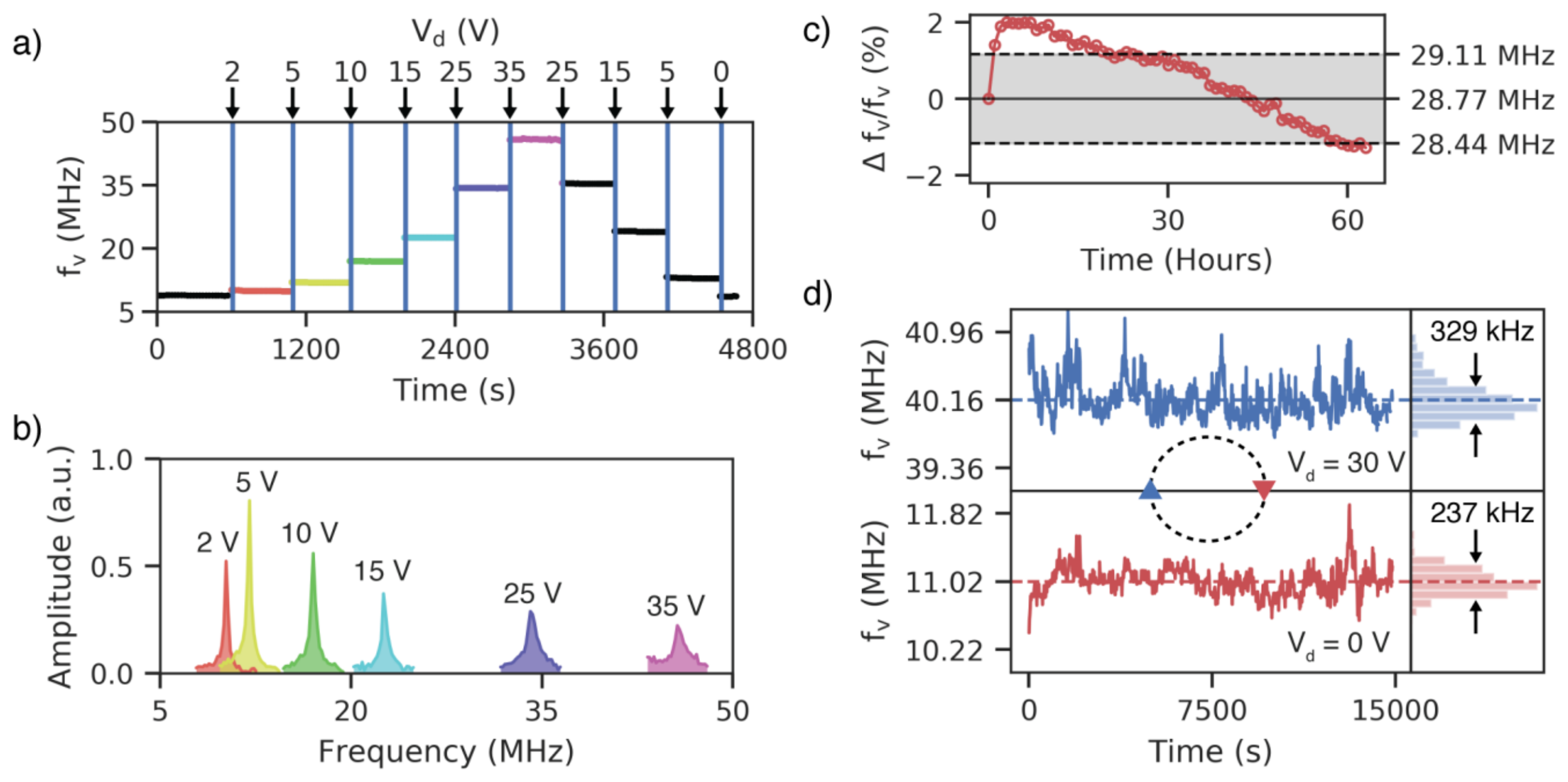

**Figure 3: a)** Setting the frequency in time. A combined high-power optical pulse of ~100 μJ and gate voltage are applied at the times indicated by arrows. **b)** Resonance spectra taken with $V_g = 0$ V immediately after photodoping for the corresponding $V_d$ in (a). The amplitude varies due to different transduction and actuation efficiencies at the different values of $V_{eff}$. **c)** Stability of $f_V$ with $V_g = 0\ V$ after phototuning to 28.77 MHz. After an initial jump of 2%, the $f_V$ decays at a rate of ~0.05%/Hour. The gray band shows the mechanical linewidth corresponding to a $Q$ of 83. **d)** Reproducibility of $f_V$ for 919 complete cycles of phototuning with $V_d = 30$ V followed by erasure with $V_d = 0$ V. The $f_V$ value falls within a ~0.3 MHz band for both states.

To demonstrate the reversibility of $f_V$ using phototuning, we change $f_V$ at discrete time intervals by varying the doping potential. At the beginning of each interval, we phototune the device using a single short, high power laser pulse (~2 mW, 0.5 s) at a given $V_d$ (Figure 3a), and then continuously monitor the $f_V$ by fitting the resonance spectra (Figure 3b) for the remainder of the interval, ~600 seconds. Increasing $V_d$ stepwise from 0 to 35 V (as seen in Figure 3a), $f_V$ takes on fixed, stable values that increase from 7 MHz up to 45 MHz. When we decrease $V_d$ stepwise back to 0 V, $f_V$ returns to 7 MHz. This data clearly demonstrates that the phototuning of $f_V$ is both reversible and bidirectional. The tuning range of $f_V$ is large, here nearly 550%, which is an order of magnitude larger than previous hybrid tuning methods[26,27]. For a quality factor of a typical device ($Q\sim100$), this tuning range equals 543 resonance linewidths. In our measurements we limited the doping potential to 35 V to avoid damage to the mechanical resonators, but larger potentials up to the dielectric breakdown of the $SiO_2$ could be used to achieve an even higher degree of tuning.

The frequency phototuning method is persistent. This persistence is clear from the steps in Figure 3a, which show $f_V$ is stable for at least 600 s. To assess the longer-term stability of phototuning, we write $f_V$ a single time and then measure $f_V$ every hour over the course of 3 days. Figure 3c plots the fractional change $\Delta f_V/f_V$ after phototuning $f_V$ with a doping potential of 30 V. Initially, $f_V$ blueshifts by 2% over the course of 2 hours. After this initial detuning, the device slowly relaxes and $f_V$ redshifts at a rate of 0.05%/hour. For reference, the mechanical linewidth for our devices is $\sim$2% of the resonance frequency (shaded region of Figure 3b), so the frequency shifts by a linewidth in $\sim$40 hours. This long-lived state does not require an external power supply or gate bias. Therefore, phototuning can replace patterned gate electrodes[13,37], even in arbitrarily large resonator arrays. To isolate the effect of the probe laser, which will cause some detuning, we set $f_V$ and measure it once after 8 days (See supporting information). We still observe a small amount of detuning in addition to warping of the gate tuning spectrum. Therefore, additional sources of detuning are present and may include the rearrangement of the trapped charge in the oxide or h-BN, or strain relaxation (*e.g.* in folds and edge clamping). Improved stability would likely be possible at cryogenic temperatures, which would reduce thermally-induced recombination of the ionized defects[32]. Given the time scale of the drift, phototuning feedback would be a straightforward means to stabilize the frequency.

The phototuning method can achieve a high degree of frequency tuning repeatability and can execute an indefinite number of write/erase cycles with no observable change to the mechanical properties of the NEMS device. To test repeatability and cycling performance, we erase the frequency state by phototuning with $V_d = 0$ V, then we write $f_V$ with $V_d = 30$ V. For all writing and erasure steps, the same dose of $\sim 2$ mW over 0.5 s was used. Figure 3c shows the results after 919 erase/write cycles. As measured from the histogram (right of Figure 3d), the average frequencies of erased and written states are $f_{erase} = 11.02 \pm 0.12$ MHz and $f_{write} = 40.16 \pm 0.16$MHz, which yields a writing repeatability of 99.5%. The small uncertainty in the repeatability could be inherent to the phototuning process, but could also be caused by sources of frequency noise and fluctuations common to 2D NEMS, such as adsorbates, heating, and unwrinkling[31], or fluctuations in the power of doping laser. The large frequency separation of the written and erased states in Figure 3a and Figure 3c could easily allow a discrete binary logic state[8,39] or, given the measured error of 160 KHz, over 150 discrete and well-defined logic states.

The temporal rate of the phototuning method is exceptionally fast. The phototuning rate can be inferred from the time-dependence of either $f_V$ or $V_{mCNP}$ during the phototuning process (see supporting information). Figure 4a shows a plot of $V_{mCNP}(t)$ (blue, upper) and $f_V(t)$ (orange, lower) for $V_d = 9$ V, and $P \sim 530$ µW with a 445 nm laser. Both $V_{mCNP}$ and $f_V$ approach steady state saturation values within $\sim 10$ ms. As noted earlier, we find that $V_{mCNP}$ does not saturate exactly to $V_d$, but each device has a small but consistent offset, which we denote $\delta V_{mCNP}$. To obtain the doping rate, we approximate $V_{mCNP}(t)$ with a saturation function of the form,

$$V_{mCNP}(t) = \Delta V(1 - e^{-\alpha t}) + V_0 \tag{1}$$

where $\Delta V \approx V_d - V_0 + \delta V_{mCNP}$, $V_0$ is fixed at the initial $V_{mCNP}$, and $\alpha$ is the doping rate, which depends on the laser's power, wavelength, and position[35,36]. Prior to each rate measurement, the device is photodoped at high power with $V_d = 0$ V, which initializes $V_0$ to $\delta V_{mNCP}$. The black trace in the upper plot of Figure 3a is the fit for $V_{mCNP}(t)$ using Eq. 1, with fit parameters $\Delta V = 8.94$ V and $\alpha = 129$ s$^{-1}$. In this model, the instantaneous doping rate is $\left|\frac{dV_{mCNP}(0)}{dt}\right| = |\Delta V|\alpha$, and the frequency tuning rate is $R_f \equiv \left|\frac{dV_{mCNP}(0)}{dt}\right| \frac{df_0}{dV_g} = \alpha|\Delta V|\frac{df_0}{dV_g}$, where $\frac{df_0}{dV_g}$ is the slope of the photodoped gate voltage tuning curve at $V_g = 0$ V (*i.e.* at $f_V$). With

$|V_{mCNP}|\sim 8$ V, $\frac{df}{dV_g}$ is between $0.2 - 1.5$ MHz/V (see supporting information), which with $|\Delta V| = 1$ V gives a range of $R_f \sim 24 - 181$ MHz/s. $\frac{df_0}{dV_g}$ is determined by the device geometry[24] and strain, and could be increased by using small area devices or shallower cavities. We note that $R_f$ characterizes the change in the steady-state $f_V$ for a particular dose, not the dynamic change in $f_V$, which is limited by the RC time constant of the device ($\sim 1$ μs). Still, the frequency tuning rate of phototuning is exceptionally fast; for example, with a moderate bias voltage of $\sim 10$ V and optical power of 1 mW, it is possible to tune a resonator by its full linewidth in $\sim 100$ μs ($10^4$ resonators per second) or to enable feedback control with a bandwidth exceeding 10 kHz.

The photodoping rate depends on the device material (graphene vs. gr/hBN), the polarity of $\Delta V$, and the optical power. We measure $\alpha$ with optical power ranging from $20 - 2540\ \mu W$ (see supporting information for full power range) with a blue doping laser (445 nm) and set $V_d = \pm 8$ V. The results for two graphene/hBN and three graphene-only devices (Figure 4b-c) show several features. First, the photodoping rate depends on whether $\Delta V < 0$ ($\alpha^-$ branch) or $\Delta V > 0$ ($\alpha^+$ branch) (Figure 4b). For all devices, $\alpha^+ > \alpha^-$, but the difference can vary greatly. For the gr/hBN devices, $\alpha^+ \sim 2\alpha^-$, while for graphene-only devices $\alpha^+ \sim 10^3 \times \alpha^-$ (Figure 4c). Second, while $\delta V_{mCNP}$ is less than 15% of $V_d$ for both branches of the gr/hBN devices, this is only true for the slower $\alpha^-$ branch of the graphene devices. For the $\alpha^+$ branch of the graphene device, $\delta V_{mCNP}$ can be as large as $\sim 50\%$ of $V_d$ (Figure 4b). Third, both $\alpha$ branches follow a non-linear dependence on the power, $\alpha(P) = \alpha_0 P^\gamma$, where $\gamma$ is between $1.4 - 1.6$ for gr/hBN devices and $1.2 - 1.7$ for the graphene devices (see supporting information). Figure 4d illustrates an example of $\alpha(P)$ for the $\alpha^+$ branch of a gr/hBN device with laser power increasing from $20$ – $640\ \mu W$, where the black line is a fit with $\gamma = 1.63$. The non-linear dependence on the power shows that the phototuning effect is not purely determined by delivered energy, but is also potentially due to local laser-induced heating of the suspended 2D sheets[28], which would lower the energy barrier between the donors (acceptors) and the conduction (valence) bands. Lastly, $\alpha$ for gr/hBN devices is greater than for graphene-only devices, regardless of branch, although the difference is significantly larger for the $\alpha^-$ branch, which can differ by a factor of $10^2 - 10^3$ (Figure 4c). The differences in the rates and the value of $\Delta V$ between the gr/hBN and graphene-only devices suggests that the gr/hBN heterostructure has a higher density of ionizable dopants and/or a lower dopant ionization energy, consistent with previous reports of electronic photodoping[32,33].

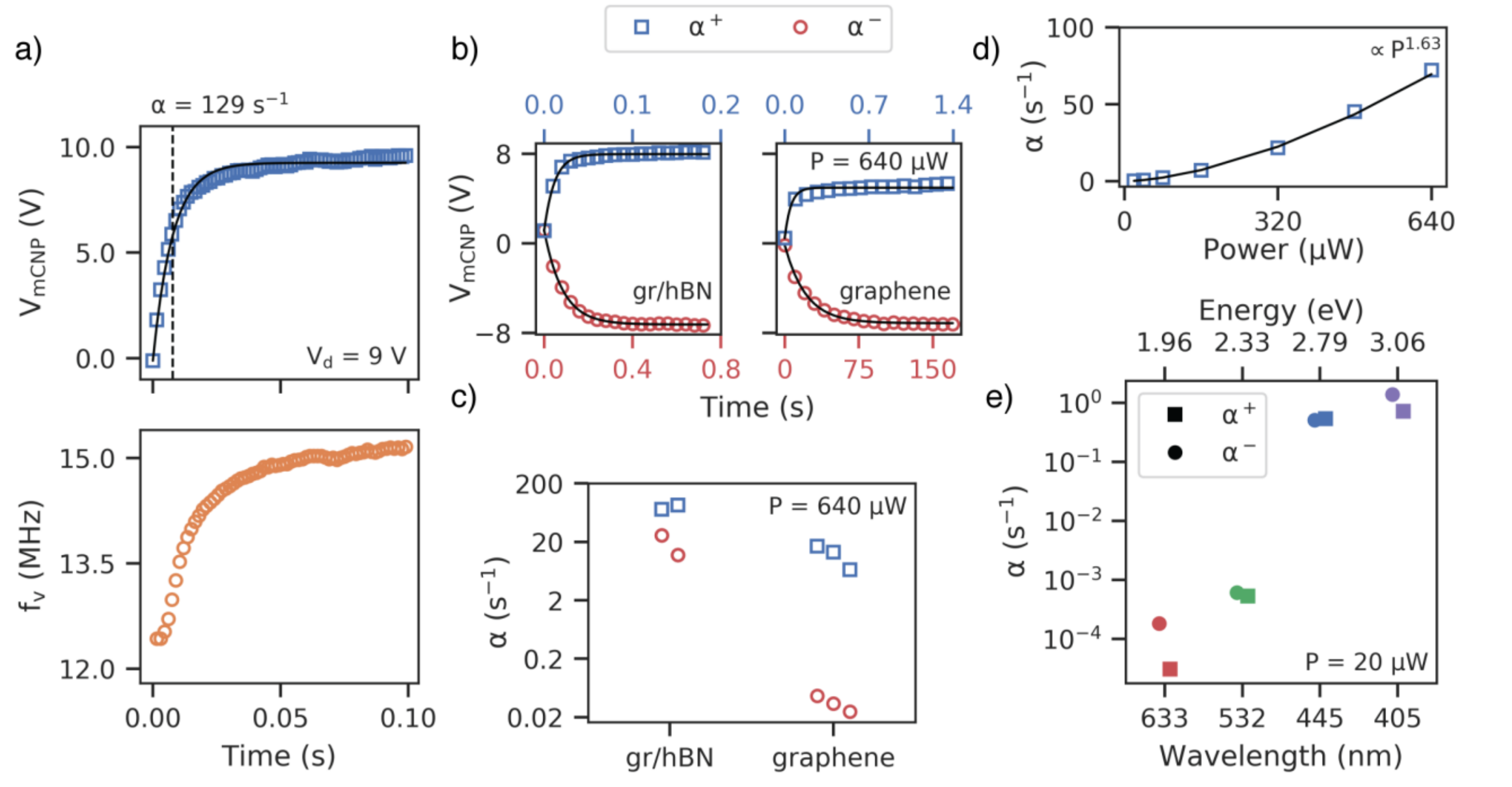

**Figure 4:** Measurement of the photodoping rate. The laser wavelength is 445 nm except in (e). **a)** Top: Example of a single photodoping curve with $V_d = 9$ V. The black line is an exponential fit to the data with $\alpha = 124$ s$^{-1}$. Bottom: Same data for $f_V$. **b)** Photodoping curves for $\Delta V > 0$ in blue ($\alpha^+$ branch) and $\Delta V < 0$ in red ($\alpha^-$ branch) for a gr/hBN device and a graphene-only device. The laser power is 640 μW and $V_d = \pm 8$ V for all the measurements. **c)** Doping rates for both $\alpha^+$ (blue squares) and $\alpha^-$ (red circles) for 640 μW, 445 nm irradiation for two gr/hBN and three graphene devices. For the $\alpha^+$ branch of the graphene devices, which don't saturate close to $V_d$, we take the initial slope of $V_{mCNP}(t)$ and divide by $V_d$ to obtain $\alpha$. **d)** Laser-power dependent doping rate for the $\alpha^+$ branch of one of the gr/hBN devices in (c). The black line is a fit to the function $\alpha = \alpha_0 P^\gamma$, with $\gamma = 1.63$. **e)** Photodoping rates of a gr/hBN device for both the $\alpha^+$ (squares) and $\alpha^-$ (circles) branches. Data was collected using 405 nm, 445 nm, 532 nm, and 633 nm illumination all at 20 μW. The photodoping rate increases by a factor of $10^4$ from 633 nm to 405 nm illumination. For the 532 nm and 633 nm laser illumination, we use the same linearized doping rate defined in (c).

The phototuning rate is greater for shorter wavelength light. To characterize the wavelength dependence of $\alpha$, we measure $\alpha$ in a gr/hBN device using four different laser wavelengths (633, 532, 445, 405 nm) with an optical power of 20 μW and $|V_d| = 8$ V. The results (Figure 4e) show that shorter wavelength, higher energy illumination leads to much faster phototuning. Compared to 633 nm light ($\alpha^+ = 1.8 \times 10^{-4}$ s$^{-1}$ and $\alpha^- = 3.1 \times 10^{-5}$ s$^{-1}$), $\alpha$ for 405 nm light ($\alpha^+ =$ 1.3 s$^{-1}$ and $\alpha^- = 0.72$ s$^{-1}$) is larger by a factor of $\sim 10^4$. The rate increase also appears to be saturating near 3 eV. The enhanced phototuning at shorter wavelengths agrees with previous photodoping studies in h-BN as well as $SiO_2$[32–34]. The wavelength dependence of $\alpha$ is advantageous for nanomechanics experiments, as it allows selection of a long-wavelength laser for transduction, which has a negligible phototuning effect, and a short-wavelength laser for phototuning. We note that higher photon energies likely also induce photodoping, which could explain the frequency shifts seen in γ-ray irradiated 2D sheets[40].

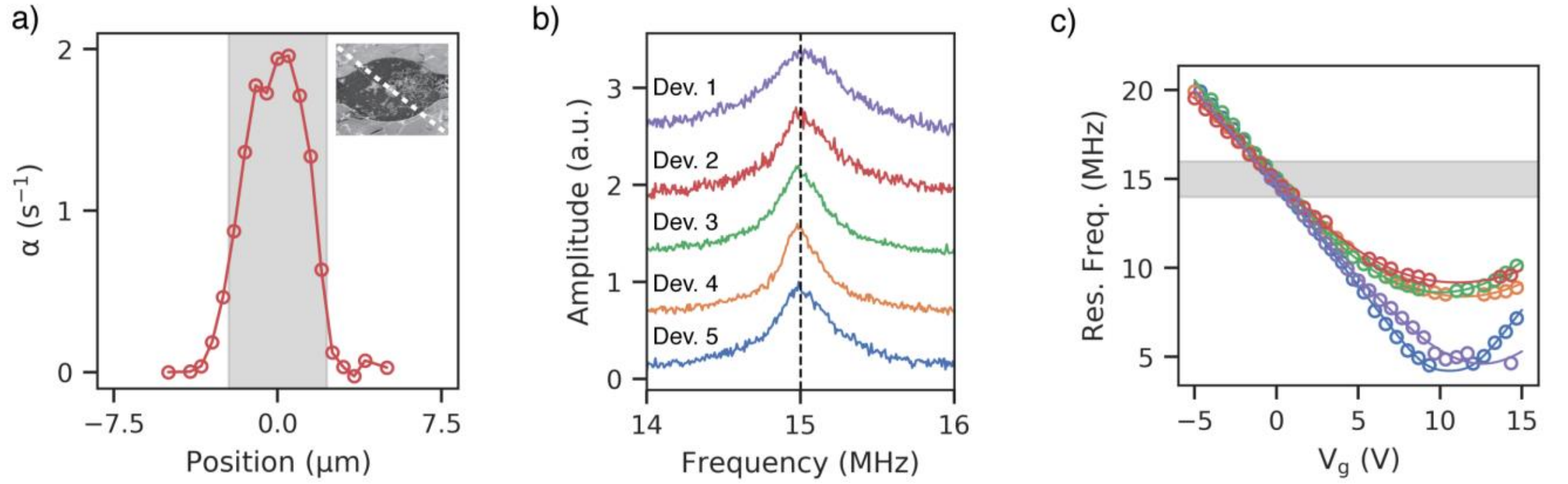

**Figure 5: a)** Photodoping rate as the doping laser is scanned across a gr/hBN drumhead. The scan line is indicated in the SEM image inset. The doping rate reaches a maximum at the center of the drum and rapidly falls off as it is swept away. The gray band indicates the spatial extent of the device. **b)** Resonant response for five different Gr/hBN devices displaying alignment of $f_V$ to 15 MHz of less than 0.2%. The alignment is achieved with no external electrical gate bias. **c)** Electrostatic tuning curves for the fundamental mode of the five devices from b (the data color corresponds to each device) after being simultaneously phototuned to $f_V = 15$ MHz. The gray horizontal band corresponds to the 14-16 MHz plot range of (b).

Many applications in NEMS circuits and lattices require precise, programmable frequency and strain tuning of individual resonators within large arrays on a single chip[13,37]. To demonstrate this capability with phototuning, we first show that the effect is localized to the laser spot. Figure 5a shows the doping rate at 20 μW measured at different locations on the membrane. The device begins to photodope only when the gaussian spot of the laser overlaps with the area of the membrane. We use the dilation of the spatial doping rate profile relative to the device diameter (greyed region in Figure 5a) to infer a spatial resolution of ∼1 μm, which is approximately the size of the laser spot. Next, we align the frequencies of five different gr/hBN devices, which are all on a single chip, to within 30 kHz (or 0.2%) of $f_V = 15$ MHz, as shown in the Figure 5b amplitude spectra, achieving a tuning precision within $\sim 5\%$ of a resonance linewidth. The resonance frequency gate curves for each device are shown in Figure 5c. While the curves intersect at $V_g = 0$ V, which defines $f_V$, the values of $V_{mCNP}$ and the general curve shape vary considerably. This demonstrates that phototuning is largely insensitive to variations between individual resonators and is thus a robust frequency tuning method.

The phototuning effect we demonstrate in graphene and gr/hBN NEMS could be applicable to other NEMS as well. Persistent photodoping has been observed in a variety of graphene heterostructures[33,34,41] as well as other materials including $SrTiO_3$[42]. In these systems, the mechanical element does not need to be graphene or an atomically thin graphene hybrid. For example, graphene-coated silicon nitride nanobeams[43] could be persistently tuned but would retain ultra-high quality factors of $Q > 10^6$. However, atomically thin resonators have the great advantage of an extreme tuning range.

Phototuning offers intriguing possibilities for both applied and fundamental physics in isolated NEMS and NEMS arrays, where tight control over individual resonators is essential. Our technique can pattern arbitrary complex geometries of static charge across a single, large-area resonator, which could improve the actuation efficiency of antisymmetric modes or allow tunable intermodal coupling[31], both commonly achieved via intricately patterned back-gates. Furthermore, this tuning also offers new opportunities for programmable NEMS crystals. In our vision for these crystals, individual resonators would be coupled to neighboring resonators by a suspended bridge material. By phototuning the resonators and the bridges, precisely tuned complex acoustic crystals would possible. Unlike previous demonstrations of static phononic crystals[5] and tunable phononic crystals[16], our approach is not vulnerable to fabrication imperfections and possesses a higher degree of tunability and the ability to modify individual unit cells of the crystal.

Finally, due to the high spatial resolution and large tuning rates, phototuning also permits high-density analog mechanical memory storage, where information is encoded in the frequency of each resonator. If each memory state is separated by a resonance linewidth ($\sim 100$ kHz), the tuning range demonstrated in Figure 3a would accommodate $\sim 500$ states, equivalent to a 9-bit memory. Given the measured tuning rates of 1 GHz/s, the state of a single device could be uniformly tuned by a linewidth ($f_V/Q \sim 100$ kHz) in $\sim 100$ μs. Even faster tuning control should be possible with higher laser power since both graphene and h-BN are stable at high temperatures[44] .

In conclusion, we have demonstrated a fast, reversible, persistent, and scalable frequency tuning method based on deterministic charge trapping, which allows for electro-optic "etch-a-sketch" patterning of strain in 2D NEMS arrays. Our phototuning technique eliminates the need for complex, lithographically defined gate electrodes used to electrostatically strain and frequency tune NEMS resonators. When applied to large NEMS lattices, this approach could enable reprogrammable phononic crystals and waveguides[16,37], or more exotic applications such as nanomechanical logic, neuromorphic computing[17,18], or the simulation of complex networks[19].

## METHODS:

*Fabrication of 2D drumheads:*

gr/h-BN mechanical drumhead resonators were fabricated by transferring the 2D sheets over an array of cavities etched into 1 μm wet thermal oxide[45,46] grown on degenerately doped silicon wafers (University Wafer). The cavities were fabricated using direct-write optical lithography and $CHF_3$ based reactive ion etching. A $\sim 300$ nm layer of oxide was left at the bottom of the cavity to act as a charge trapping layer and to prevent shorting. Ti/Pt electrodes were defined by lithography and deposited by electron-beam evaporation.

To prepare the 2D sheets for transfer, a relatively thick layer ($\sim 3$ μm) of PMMA A11 was spun onto CVD grown single-layer h-BN on Cu foil (Graphene Supermarket) and then a polyamide scaffold with a central hole removed was then placed on the PMMA/hBN/Cu stack. The stack was placed in a bath of Ammonium Persulphate to etch the Cu and then rinsed in deionized water and dried in air. The polyamide/PMMA/hBN was placed on top of CVD graphene grown on Cu foil (Graphenea) and baked at 180 °C for 30 minutes to adhere the hBN and the graphene[46]. The etching, rinsing, and drying was repeated leaving a freestanding film of PMMA/hBN/Graphene supported by the polyamide scaffold. To transfer the 2D sheets to the cavity substrates, the PMMA/hBN/Graphene stack was then placed graphene-side-down on top of the pre-patterned cavities and adhered at 155 °C overnight (∼15 hours). After removing the polyamide scaffold, the PMMA was removed in flowing $Ar/H_2$ at 400 °C. The graphene sheet contacts the electrodes from above, resulting in a simultaneous electrical connection to all devices. Graphene-only devices were fabricated in a similar fashion with both an in-house and a commercial transfer process performed by Graphenea.

*Measurement of mechanical motion:*

Device motion was measured using optical interferometry, as described previously[28]. A 633 nm HeNe laser was focused onto the devices (held at room temperature at $10^{-6}$ torr) using a 40 ×, 0.6 NA objective. The reflected light was detected using a high-sensitivity photodiode (Thorlabs APD 130A) and the voltage signal was demodulated using a Zurich Instruments HFLI2 Lock-In amplifier. The incident laser was scanned with a two-axis galvometer and passed through an optical relay system in order to image the mode shape and to maximize transduction sensitivity. We used low laser power (∼1 – 10 µW) to avoid unwanted photodoping by the 633 nm probe laser.

*Photodoping:*

A separate laser (405 nm, 445 nm, or 532 nm) was used for photodoping. The doping laser was coupled into the beam-path using a dichroic mirror and focused onto the sample using the same 40 ×, 0.6 NA objective lens. A separate two-axis galvometer was used to position the doping laser at the center of the drumheads. The laser power for each color was calibrated using a power meter (Thorlabs 120VC) and maintained using PID control. For dynamic measurements of $V_{mCNP}$, an acousto-optic modulator (AA-Optoelectronics MT350-A0.12-VIS) was used to supply a well-defined pulse of the doping laser with pulse-widths down to ∼10 ns. Prior to all measurements, the doping laser was scanned across the device with $V_d = 0$ V to guarantee a uniformly doped initial erased state.

SUPPORTING INFORMATION:

Method for measuring $V_{mCNP}$, diagram of optical setup, long-term stability of phototuning, measurement of $\frac{df_0}{dV_g}$, and doping rate as function of laser power.

AUTHOR INFORMATION:

Corresponding Author:

Benjamín Alemán

*baleman@uoregon.edu

AUTHOR CONTRIBUTIONS:

DM and BA conceived and designed the experiments. DM and AB fabricated the NEMS devices. DM designed and built optical measurement apparatus and performed the experiments with assistance from AB. DM analyzed the data. DM and BA wrote the manuscript, and AB provided comments on the manuscript. BA supervised the work.

ACKNOWLEDGMENTS:

We acknowledge the facilities and staff from the Center for Advanced Materials in Oregon (CAMCOR), and the use of the University of Oregon's Rapid Materials Prototyping facility, funded by the Murdock Charitable Trust. We thank Joshua Ziegler, Rudy Resch, and Kara Zappitelli for scientific discussions and feedback related to this work. This work was supported by the University of Oregon and the National Science Foundation (NSF) under grant No. DMR-1532225.

REFERENCES:

(1) Jensen, K.; Kim, K.; Zettl, A. *Nat. Nanotechnol.* **2008**, *3* (9), 533–537.

(2) Steele, G. A.; Huttel, A. K.; Witkamp, B.; Poot, M.; Meerwaldt, H. B.; Kouwenhoven, L. P.; van

der Zant, H. S. J. *Science* **2009**, *325* (5944), 1103–1107.

(3) Degen, C. L.; Poggio, M.; Mamin, H. J.; Rettner, C. T.; Rugar, D. *Proc Natl Acad Sci U S A* **2009**, *106* (5), 1313–1317.

(4) Okamoto, H.; Gourgout, A.; Chang, C.-Y.; Onomitsu, K.; Mahboob, I.; Chang, E. Y.; Yamaguchi, H. *Nat. Phys.* **2013**, *9* (8), 480–484.

(5) Chan, J.; Alegre, T. P. M.; Safavi-Naeini, A. H.; Hill, J. T.; Krause, A.; Gröblacher, S.; Aspelmeyer, M.; Painter, O. *Nature* **2011**, *478* (7367), 89–92.

(6) Dominguez-Medina, S.; Fostner, S.; Defoort, M.; Sansa, M.; Stark, A.; Halim, M. A.; Vernhes, E.; Gely, M.; Masselon, C.; Hentz, S. *Science* **2018**, *362* (November), 918–922.

(7) Ekinci, K. L.; Roukes, M. L. *Rev. Sci. Instrum.* **2005**, *76* (6), 061101.

(8) Yao, A.; Hikihara, T. *Appl. Phys. Lett.* **2014**, *105* (12), 123104.

(9) Rueckes, T., Kyoungha Kim, Ernesto Joselevich, Greg Y. Tseng, Chin-Li Cheung, C. M. L. *Science* **2000**, *289* (5476), 94–97.

(10) Masmanidis, S.; Masmanidis, S.; Karabalin, R.; Vlaminck, I.; Borghs, G.; Freeman, M.; Roukes, M. *Science* **2007**, *317* (5839), 780–783.

(11) Hafiz, M. A. A.; Kosuru, L.; Younis, M. I. *J. Appl. Phys.* **2016**, *120* (7).

(12) Roukes, M. L. *IEDM Tech. Dig.* **2004**, No. December 13-15, 539–542.

(13) Cha, J.; Daraio, C. *Nat. Nanotechnol.* **2018**, *13* (11), 1016–1020.

(14) Olsson, R. H.; El-Kady, I. *Meas. Sci. Technol.* **2009**, *20* (1).

(15) Wang, Y.; Yousefzadeh, B.; Chen, H.; Nassar, H.; Huang, G.; Daraio, C. *Phys. Rev. Lett.* **2018**, *121* (19), 194301.

(16) Cha, J.; Kim, K. W.; Daraio, C. *Nature* **2018**, *564* (7735), 229–233.

(17) Hoppensteadt, F. C.; Izhikevich, E. M. *IEEE Trans. Circuits Syst. I Fundam. Theory Appl.* **2001**, *48* (2), 133–138.

(18) Kumar, A.; Mohanty, P. *Sci. Rep.* **2017**, *7* (1), 1–9.

(19) Matheny, M. H.; Emenheiser, J.; Fon, W.; Chapman, A.; Salova, A.; Rohden, M.; Li, J.; Hudoba de Badyn, M.; Pósfai, M.; Duenas-Osorio, L.; Mesbahi, M.; Crutchfield, J. P.; Cross, M. C.; D'Souza, R. M.; Roukes, M. L. *Science* **2019**, *363* (6431), eaav7932.

(20) Zhang, W.-M.; Hu, K.-M.; Peng, Z.-K.; Meng, G. *Sensors (Basel).* **2015**, *15* (10), 26478–26566.

(21) Bunch, J. S.; Van Der Zande, A. M.; Verbridge, S. S.; Frank, I. W.; Tanenbaum, D. M.; Parpia, J. M.; Craighead, H. G.; McEuen, P. L. *Science* **2007**, *315* (January), 490–493.

(22) Merced, E.; Cabrera, R.; Dávila, N.; Fernández, F. E.; Sepúlveda, N. *Smart Mater. Struct.* **2012**, *21* (3).

(23) Chiao, M.; Lin, L. *J. Micromechanics Microengineering* **2004**, *14* (12), 1742–1747.

(24) Chen, C.; Rosenblatt, S.; Bolotin, K. I.; Kalb, W.; Kim, P.; Kymissis, I.; Stormer, H. L.; Heinz, T. F.; Hone, J. *Nat. Nanotechnol.* **2009**, *4* (12), 861–867.

(25) Enderling, S.; Hedley, J.; Jiang, L.; Cheung, R.; Zorman, C.; Mehregany, M.; Walton, A. J. *J. Micromechanics Microengineering* **2007**, *17* (2), 213–219.

(26) Kim, K.; Jensen, K.; Zettl, A. *Nano Lett.* **2009**, *9* (9), 3209–3213.

(27) Chang, J.; Koh, K.; Min, B. K.; Lee, S. J.; Kim, J.; Lin, L. *ACS Appl. Mater. Interfaces* **2013**, *5* (19), 9684–9690.

(28) Davidovikj, D.; Slim, J. J.; Cartamil-Bueno, S. J.; van der Zant, H. S. J.; Steeneken, P. G.; Venstra, W. J. *Nano Lett.* **2016**, *16* (4), 2768–2773.

(29) Song, X.; Oksanen, M.; Sillanpää, M. A.; Craighead, H. G.; Parpia, J. M.; Hakonen, P. J. *Nano Lett.* **2012**, *12* (1), 198–202.

(30) Will, M.; Hamer, M.; Müller, M.; Noury, A.; Weber, P.; Bachtold, A.; Gorbachev, R. V.; Stampfer, C.; Güttinger, J. *Nano Lett.* **2017**, *17* (10), 5950–5955.

(31) De Alba, R.; Massel, F.; Storch, I. R.; Abhilash, T. S.; Hui, A.; McEuen, P. L.; Craighead, H. G.; Parpia, J. M. *Nat. Nanotechnol.* **2016**, *11* (9), 741–746.

(32) Ju, L.; Velasco, J.; Huang, E.; Kahn, S.; Nosiglia, C.; Tsai, H.-Z.; Yang, W.; Taniguchi, T.; Watanabe, K.; Zhang, Y.; Zhang, G.; Crommie, M.; Zettl, a; Wang, F. *Nat. Nanotechnol.* **2014**, *9* (May).

(33) Kim, Y. D.; Bae, M.-H.; Seo, J.-T.; Kim, Y. S.; Kim, H.; Lee, J. H.; Ahn, J. R.; Lee, S. W.; Chun, S.-H.; Park, Y. D. *ACS Nano* **2013**, *7* (7), 5850–5857.

(34) Choi, H. H.; Park, J.; Huh, S.; Lee, S. K.; Moon, B.; Han, S. W.; Hwang, C.; Cho, K. *ACS Photonics* **2018**, *5* (2), 329–336.

(35) Neumann, C.; Rizzi, L.; Reichardt, S.; Terrés, B.; Khodkov, T.; Watanabe, K.; Taniguchi, T.; Beschoten, B.; Stampfer, C. *ACS Appl. Mater. Interfaces* **2016**, *8* (14), 9377–9383.

(36) Velasco, J.; Ju, L.; Wong, D.; Kahn, S.; Lee, J.; Tsai, H. Z.; Germany, C.; Wickenburg, S.; Lu, J.; Taniguchi, T.; Watanabe, K.; Zettl, A.; Wang, F.; Crommie, M. F. *Nano Lett.* **2016**, *16* (3), 1620–1625.

(37) Hatanaka, D.; Bachtold, A.; Yamaguchi, H. *Phys. Rev. Appl.* **2019**, *11* (2), 024024.

(38) Schedin, F.; Geim, A. K.; Morozov, S. V.; Hill, E. W.; Blake, P.; Katsnelson, M. I.; Novoselov, K. S. *Nat. Mater.* **2007**, *6* (9), 652–655.

(39) Mahboob, I.; Yamaguchi, H. *Nat. Nanotechnol.* **2008**, *3* (5), 275–279.

(40) Lee, J.; Krupcale, M. J.; Feng, P. X. L. *Appl. Phys. Lett.* **2016**, *108* (2).

(41) Ho, P.-H.; Chen, C.-H.; Shih, F.-Y.; Chang, Y.-R.; Li, S.-S.; Wang, W.-H.; Shih, M.-C.; Chen, W.-T.; Chiu, Y.-P.; Li, M.-K.; Shih, Y.-S.; Chen, C.-W. *Adv. Mater.* **2015**, *27* (47), 7809–7815.

(42) Yeats, A. L.; Pan, Y.; Richardella, A.; Mintun, P. J.; Samarth, N.; Awschalom, D. D. *Sci. Adv.* **2015**, *1* (9), e1500640.

(43) Verbridge, S. S.; Parpia, J. M.; Reichenbach, R. B.; Bellan, L. M.; Craighead, H. G. *J. Appl. Phys.* **2006**, *99* (12).

(44) Ye, F.; Lee, J.; Feng, P. X.-L. *Nano Lett.* **2018**, *18* (3), 1678–1685.

(45) Suk, J. W.; Kitt, A.; Magnuson, C. W.; Hao, Y.; Ahmed, S.; An, J.; Swan, A. K.; Goldberg, B. B.; Ruoff, R. S. *ACS Nano* **2011**, *5* (9), 6916–6924.

(46) Shautsova, V.; Gilbertson, A. M.; Black, N. C. G.; Maier, S. A.; Cohen, L. F. *Sci. Rep.* **2016**, *6* (1), 30210.

TABLE OF CONTENTS GRAPHIC:

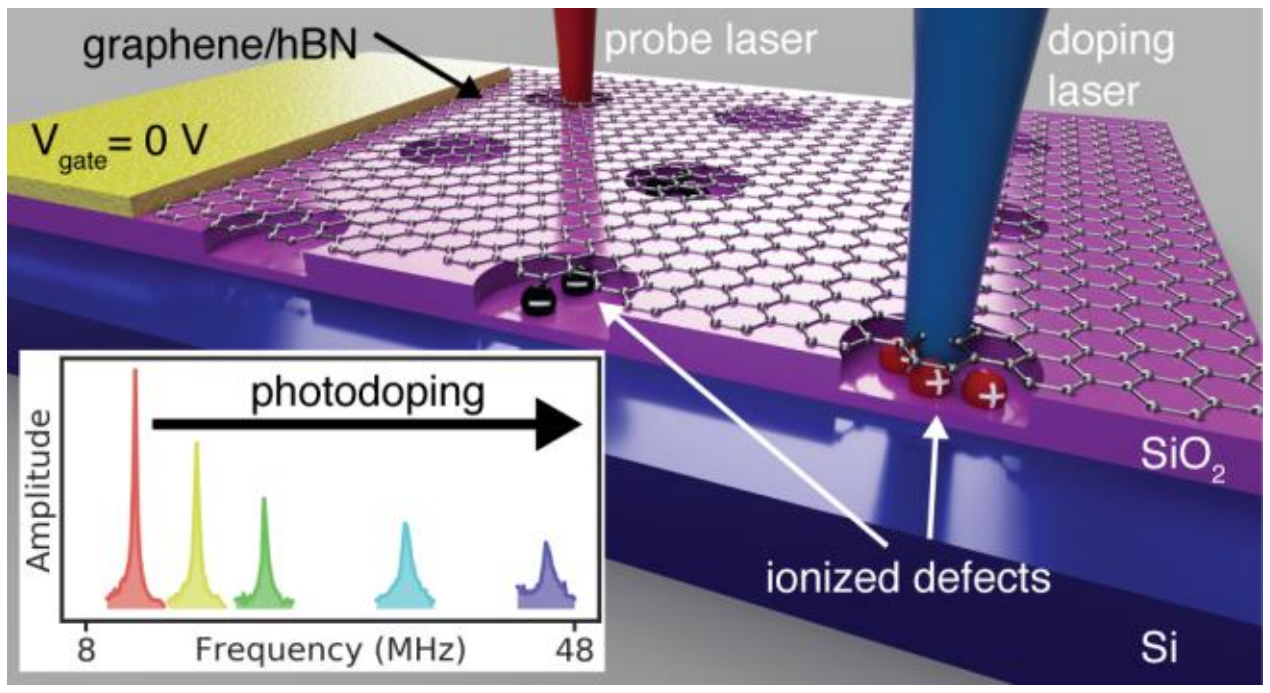